\documentclass[runningheads]{llncs}
\usepackage[T1]{fontenc}
\usepackage{algorithmic}
\usepackage{graphicx}
\usepackage{textcomp}
\usepackage{xcolor}
\usepackage{url}
\usepackage{multirow}
\usepackage{subcaption}
\usepackage{mwe}
\usepackage{cprotect}
\usepackage{adjustbox}
\usepackage{cite}

\usepackage{listings}

\usepackage{tcolorbox}
\definecolor{codegreen}{rgb}{0,0.6,0}
\definecolor{codegray}{rgb}{0.5,0.5,0.5}
\definecolor{codepurple}{rgb}{0.58,0,0.82}
\definecolor{backcolour}{RGB}{252, 253, 246}
\lstdefinestyle{mystyle}{
    backgroundcolor=\color{backcolour},   
    commentstyle=\color{codegreen},
    keywordstyle=\color{magenta},
    numberstyle=\tiny\color{codegray},
    stringstyle=\color{codepurple},
    basicstyle=\ttfamily\footnotesize,
    breakatwhitespace=true,
    breaklines=true,
    captionpos=b,
    keepspaces=true,
    numbers=left,
    numbersep=3pt,
    showspaces=false,
    showstringspaces=false,
    showtabs=false,
    tabsize=2,
    breakautoindent=false,
    breakindent=0ex,
    xleftmargin=0.25cm,frame=lr,framesep=8pt,framerule=0pt
}
\usepackage{caption}
\newcounter{promptcounter}

\newcommand{\promptcaption}[1]{%
    \refstepcounter{promptcounter}%
    \raggedright Prompt \thepromptcounter: #1
}
\begin{document}
\title{Accelerating End-to-End PDF to Markdown Conversion through Assisted Generation}
\titlerunning{Copy Lookup Decoding}
%
\author{Changxu Duan\orcidID{0000-0003-0547-0901}}
\authorrunning{C. Duan}
%
\institute{Technische Universität Darmstadt \\ \email{duan@linglit.tu-darmstadt.de}}
\maketitle
\begin{abstract}
Converting data from machine-unreadable formats like PDFs into Markdown has the potential to enhance the accessibility of scientific research.
Existing end-to-end decoder transformer models can transform screenshots of PDFs into Markdown, offering more flexibility than pipeline-based methods. 
Yet, decoding text token by token from scratch is inefficient, especially when dense text can be directly copied from the PDF.
To address this challenge, this paper modifies Prompt Lookup Decoding (PLD) to extract candidate sequences directly from PDF files, leveraging the high n-gram overlap between PDFs and their Markdown equivalents. 
A new method, Copy Lookup Decoding (CLD), is introduced here to enhance PLD’s candidate generation mechanism. Experiments demonstrate that CLD can accelerate the conversion process by up to 1.70$\times$ at original quality.
The codebase for this paper is open-source on GitHub \footnote{\url{https://github.com/Fireblossom/CopyLookup}}.

\end{abstract}
\section{Introduction}
Scientific research is almost exclusively published in PDF, which is an unstructured text format whose content is not readily machine-readable. PDF formatted scientific documents contain complex elements like mathematical formulas, figures, headers, and tables, as well as densely laid-out text.
Vision-Language Models (VLM) can understand documents in the form of document screenshots, but their comprehension is not yet reliable enough \cite{fu2024isobench}.

Markdown is an ideal format in terms of machine-readability due to its simple syntax and structured format which preserves context and enhances comprehensibility. Its support for rich text elements such as boldface, italics, and hyperlinks enriches the data, providing a more detailed context for models to generate accurate outputs. Furthermore, the Markdown community has extended the syntax so that it can support tables as well as advanced mathematical and chemical formulas \cite{mathpix-markdown-it}.
Therefore, converting PDF files to Markdown is worth the effort.

\begin{figure}
    \centering
    \includegraphics[width=0.8\linewidth]{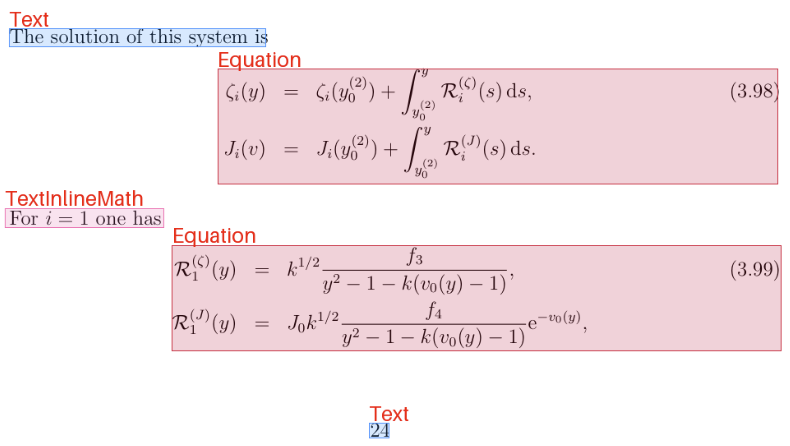}
    \caption{Example of Marker’s layout analysis on a scientific document with a complex layout. Marker detects math equations and plain text lines correctly. However, the inline math expression and the plain text line near it were incorrectly merged; the page number was also recognized as plain text. Error results from the document layout analysis accumulate and are propagated to the next steps in the pipeline.}
    \label{fig:error_example}
\end{figure}

A widely used method for converting PDFs to Markdown utilizes a pipeline-based approach, as demonstrated by Marker \cite{marker}. This process begins with a document layout analysis that segments each PDF page into various blocks and identifies the type of each block. Marker then determines the reading order of these blocks. Based on their types, different modules are called to convert specific elements, such as formulas and tables, into Markdown. The process concludes with a cleanup phase to correct errors. However, pipeline-based approaches can struggle with documents that have complex layouts, as errors made in one step can propagate and accumulate through subsequent steps, impacting overall performance, as shown in Figure~\ref{fig:error_example}.

An alternative to pipeline-based methods is end-to-end decoder transformer models, such as Nougat \cite{blecher2023nougat}. 
The model is a tiny VLM that reads a screenshot of a PDF page and directly extracts the content into Markdown, automatically preserving the correct reading order and converting mathematical formulas into \LaTeX{} code. 
However, Nougat faces limitations in processing speed as it generates text token-by-token from scratch, which can significantly slow down the document conversion process.

Assisted Generation \cite{gante2023assisted, speculativedecoding} can accelerate transformer decoder models without degrading generation quality because all candidates are validated by the transformer model.
Prompt Lookup Decoding (PLD) \cite{saxena2023prompt} proves particularly advantageous when there is a substantial overlap of n-grams between the input (prompt) and the output of LLM. This method is up to 2.4$\times$ faster in tasks such as summarization and contextual question answering.
Given that PDFs frequently contain dense text suitable for direct copying, implementing PLD could streamline the conversion process and decrease computational expenses.

The contributions of this paper include:
\begin{enumerate}
\item Modifying original PLD (mPLD) to enable the extraction of candidate sequences from PDF files. mPLD is plug-and-play, requires no fine-tuning, and is compatible with any end-to-end PDF-to-Markdown conversion models.
\item Proposing Copy Lookup Decoding (CLD) to enhance candidate generation in mPLD.
\item Releasing the dataset-making scripts and the arXiv IDs of the documents used in the experiments to enhance reproducibility and observe copyright.
\end{enumerate}

\section{Related Work}
\subsection{Pipeline-based PDF-to-Markdown Conversion}
GROBID \cite{GROBID} is a machine learning library for extracting, parsing, and restructuring documents, including PDFs, into structured XML encoded documents, which can be converted to Markdown. However, it is not flexible because it converts formulas and tables into images, thus hampering subsequent accessibility. DocTR \cite{doctr2021} and DocBed \cite{DocBed} first identify the document layout and then extract text content. Marker, MinerU \cite{2024mineru} and Docling \cite{Docling} realize PDF-to-Markdown conversion through text layout analysis, reading order detection, formula or table extraction, and finally combine them into Markdown. Some close-sourced online services such as LlamaParse\footnote{\url{https://cloud.llamaindex.ai/parse}} and Mistral OCR \footnote{\url{https://mistral.ai/news/mistral-ocr}} also provide PDF-to-Markdown conversion, using a process similar to Marker as described in their documentation.

\subsection{End-to-End PDF-to-Markdown Conversion}
End-to-end PDF-to-Markdown conversion models build on top of Vision-Language Models (VLM) and Document Understanding (DU) transformer models. Donut \cite{kim2022donut} is a DU model consisting of a visual encoder and a language model decoder without obtaining texts directly from the document. Donut's pre-training objective is to retell the text from a document screenshot.
Nougat \cite{blecher2023nougat} follows Donut in implementing a screenshot-to-Markdown conversion of academic documents with an arXiv-sourced dataset. 
$\mu$gat \cite{quattrini2024gat} extends Nougat’s input to handle elements that cross the single-page limit.
LOCR \cite{sun-etal-2024-locr} solves the problem of Nougat’s hallucination and repetition using an additional location prompt. 
Kosmos-2.5 \cite{lv2023kosmos} implements a generalized screenshot-to-Markdown conversion with a larger model size. 
The approaches described in this paper attempt to speed up Nougat and Kosmos-2.5.

General-purpose VLMs are also used to achieve screenshot-to-Markdown conversions. LlamaOCR \cite{llama-ocr}, Zerox OCR \cite{zerox} and Parallex \cite{parallex} all feed screenshots of PDFs to open- and closed-source VLMs and direct the models to return the desired output by passing in the appropriate prompts.
In this paper we replicate their methodology and conduct experiments on acceleration with two off-shelf open-source models Llama3.2\cite{grattafiori2024llama3herdmodels} and Qwen2-VL\cite{wang2024qwen2vlenhancingvisionlanguagemodels} as backbones.

\subsection{Document Layout Analysis}
Recent Document Layout Analysis (DLA) models have become increasingly powerful thanks to the availability of large-scale document layout datasets \cite{PubLayNet, li-etal-2020-docbank, DocLayNet, FUNSD}. Computer vision models have been able to extract layouts in screenshots of documents \cite{9412557, 9156933, 9428465}. Language models have also been applied to recognize layouts. LayoutLM \cite{layoutlm} and its variant VILA \cite{shen-etal-2022-vila} are transformer encoder models that analyze document layouts from the texts and their 2D coordinates. LayoutLMv2 and v3 \cite{xu-etal-2021-layoutlmv2, LayoutLMv3} additionally attach visual features to the transformer encoder.

In this work, our Copyable Text Identification module for CLD is fine-tuned from ERNIE Layout \cite{peng-etal-2022-ernie}.

\subsection{Assisted Generation}
To mitigate the high inference latency stemming from autoregressive decoding in Large Language Models (LLMs), Assisted Generation \cite{speculativedecoding} has emerged as a novel decoding paradigm for LLM inference. In each decoding step, this method first uses a smaller LLM drafts several future tokens efficiently, and then verifies them in parallel.
PLD \cite{saxena2023prompt} and LLMA \cite{yang2023inferencereferencelosslessacceleration} matches the last few tokens against an earlier position in the input prompt and selects the text span as a candidate. LookaheadDecoding \cite{pmlr-v235-fu24a} introduces multiple special tokens at the end of the input prompt to enable parallel drafting and convert the draft into an n-gram candidate. 
Lookahead \cite{lookahead} implements a multi-branch drafting strategy.
Due to the low time overhead of the drafting phase, these methods can consistently improve the inference speed of the target LLM.

Some recent studies \cite{li-etal-2024-eagle,pmlr-v235-li24bt,pmlr-v235-cai24b} have used the hidden states of the LLM as inputs to their draft models. However, these methods require additional parameters and need to be trained for each backbone model.
In this work, we modified PLD to demonstrate the effectiveness of our approach as PLD is the simplest to implement and has already been integrated into the huggingface library.

\section{Methodology}
Initially, PLD checks for a matching n-gram to the current prediction in the input (prompt) of the transformer model, copying several tokens directly from the prompt as candidates. Subsequently, the LLM would verify all the candidate tokens simultaneously, determining whether each is a plausible next token in sequence from left to right, until encountering the first implausible candidate. 
The transformer model would then accept the last sequence that was deemed reasonable. Modified PLD (mPLD) involves a straightforward change: shifting the source of the candidate sequences from the prompt to all the text on the PDF page. 

However, this modification introduces two significant issues. First, not all text in a PDF is readily copyable, which can limit the effectiveness of the extraction. Second, by default, mPLD extracts the candidate sequence from the first matching n-gram position on the page, posing the risk that candidates located later in the document may not be correctly matched.

To solve these problems, we designed Copy Lookup Decoding (CLD), which both filters out possible PDF noise and matches candidates to the back of the page through Copyable Text Identification and Candidate Generation.

\subsection{Copyable Text Identification}
\label{sec:ident}
Inspired by DLA-related works, we assert that whether the text is copyable or not is correlated with its layout information. Specifically, we suggest that:
\begin{enumerate}
    \item Plain text found in paragraphs should be copyable.
    \item Page elements such as mathematical formulas, tables, and figures should \textbf{not} be copyable.
    \item Elements that do not convey relevant content, including headers, footers, and page numbers, should \textbf{not} be copyable.
\end{enumerate}

Following VILA \cite{shen-etal-2022-vila} , we assume that text copyability is homogeneous at the span level. We use PyMuPDF \cite{pymupdf} to extract span-level text and bounding boxes from the PDF. PyMuPDF generally preserves the correct reading order, except in cases of floating elements such as footnotes and captions.
We fine-tune the ERNIE Layout model for token classification using LoRA \cite{hu2022lora} and an unofficial implementation \cite{ernie-layout-pytorch} that supports an extended input length of 1024 tokens through RoPE \cite{su2023roformerenhancedtransformerrotary}, addressing the issue of PDF pages often exceeding 512 tokens. 
The classification head of ERNIE Layout was modified to classify tokens as either \verb|KEEP| or \verb|DELETE|, simplifying the target from the broader range of layout categories typically used in DLA models (e.g., up to 10 categories in VILA) to just two.

\begin{figure}
    \centering
    \includegraphics[width=0.8\linewidth]{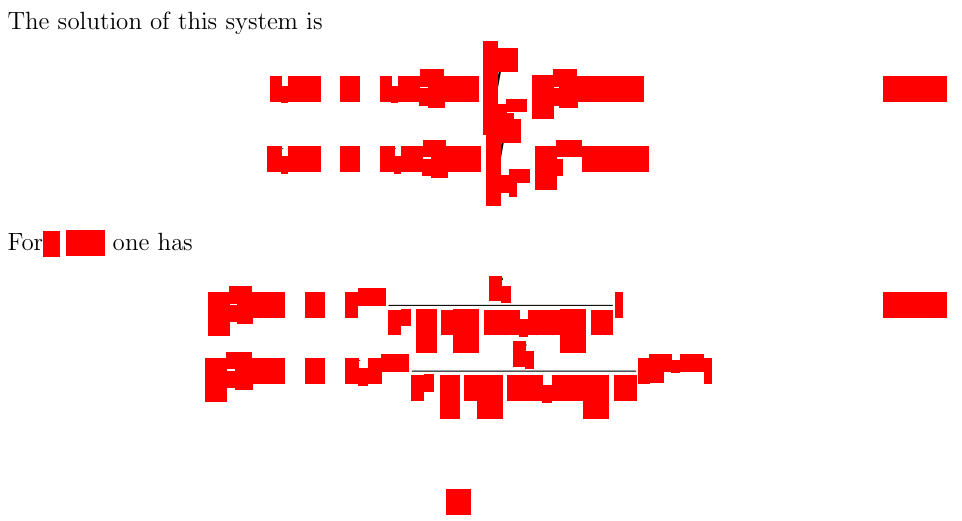}
    \cprotect\caption{Example of Copyable Text Identification. In this example, our model detects the same scientific document as in Figure~\ref{fig:error_example}. The \verb|DELETE| parts, including in-line and standalone mathematical formulas, and page numbers are identified and masked out. The \verb|KEEP| parts including copyable texts are returned by the CLD's candidate generator.}
    \label{fig:cti_example}
\end{figure}

As we have fine-tuned ERNIE Layout for token classification, and each span may contain more than one token, a voting classifier \cite{6065432} is applied to decide the prediction of the spans. 
We demonstrate an example of Copyable Text Identification in Figure~\ref{fig:cti_example}.

\subsection{Candidates Generation}
After identifying which spans of text in the PDF are copyable, we merge adjacent copyable spans while keeping non-adjacent copyable spans separate. All merged spans are stored in a list for subsequent processing.

During candidate generation, CLD checks each span in the list, starting from the first, to find a matching n-gram. If no match is found in the first span, the search continues through subsequent spans until a match is found or all the spans are exhausted. When a matching n-gram is detected in any span, the following tokens from that span are taken as candidates.

To align with the PDF-to-Markdown conversion's adherence to reading order, we implemented a topping mechanism. 
Once candidates are extracted, the PDF-to-Markdown conversion model evaluates whether these candidates are acceptable tokens in sequence. 
If a candidate is accepted, it indicates that the model is progressing within the span containing that candidate. 
At this juncture, we reassemble the span list by moving the currently active span and all subsequent spans to the front of the list while relocating all preceding spans to the end. 
This approach allows CLD to dynamically generate candidates that are synchronized with the ongoing progress of the PDF-to-Markdown conversion, enhancing both candidate accuracy and efficiency.

\section{Experiments}
\subsection{PDF-to-Markdown Models}
We test the conversion times and quality differences for scratch generation, mPLD, and CLD using two publicly available PDF-to-Markdown conversion specialized models, plus two general-purpose and open-sourced VLMs across the three test datasets described in Section~\ref{sec:dataset}. 
All tests were conducted on a machine with 1$\times$A100 GPU.

\begin{itemize}
    \item \textbf{Task specialized models: }Nougat\footnote{\url{https://huggingface.co/facebook/nougat-base}} and Kosmos-2.5\footnote{\url{https://huggingface.co/microsoft/kosmos-2.5}}
    \item \textbf{General-purpose VLMs: } Llama-3.2-11B-Vision-Instruct\footnote{\url{https://huggingface.co/meta-llama/Llama-3.2-11B-Vision}} and Qwen2-VL-72B-Instruct-AWQ\footnote{\url{https://huggingface.co/Qwen/Qwen2-VL-72B-Instruct-AWQ}}. Qwen2-VL is labeled as a 12.6B parameter in the model card because of the quantization technique \cite{lin2023awq}. We use Prompt in Prompt~\ref{prompt} for these models.
\end{itemize}

\begin{figure}[!thp]
\centering
\begin{minipage}{0.96\linewidth}
\begin{lstlisting}
Convert the following PDF page to Markdown. Return only the Markdown with no explanation text.
Leave out any page numbers and redundant headers or footers.
Do not include any code blocks (e.g. "```markdown" or "```") in the response. If unable to parse, return an empty string.
\end{lstlisting}
\end{minipage} 
\promptcaption{Prompt that instructs Llama and Qwen2-VL to complete PDF-to-Markdown conversion task.}
\label{prompt} 
\end{figure}

\subsection{Dataset Building}
\label{sec:dataset}
As there is no existing dataset released as PDFs at this time, we downloaded the \LaTeX{} source code bundles for the July and August 2023 papers from arXiv. Then we used \LaTeX-Rainbow \cite{duan-etal-2023-latex}, which compiles \LaTeX{} to PDF and annotates semantic labels, reading order, and \LaTeX{} code corresponding to mathematical formulas and tables for each element on a page. A total of 14,320 papers were annotated.

Spans are extracted from these pages and are labeled as either \verb|KEEP| or \verb|DELETE| based on the results of the annotation from the previous step. Pages that are challenging to read, such as those containing full-page images, long tables, or bibliographies, are excluded from the dataset. Finally, a dataset was assembled consisting of 180,146 pages, each annotated with span-level text copyable labels and their corresponding bounding boxes. We randomly split the training set to 162,127 pages and the test set to 18,019 pages. The vast majority of the pages in this dataset are in English.

Our method extracts text spans from PDFs, which requires access to the full text of academic papers. As arXiv does not grant permission to repost the full text\footnote{\url{https://info.arxiv.org/help/license/reuse.html\#full_text}}, we publish the scripts for creating the datasets along with the arXiv numbers of the documents to provide reproducibility.

For the PDF-to-Markdown conversion speed test, we randomly selected 128 pages from the test set. In addition, we also randomly selected 128 pages each from papers categorized in Economics and Quantum Physics respectively to test the performance of mPLD and CLD in different domain documents.

\section{Results}
\subsection{Copyable Text Identification}
We use F1 scores as the metric, without averaging because our model is a binary classifier. We tested token-level scores and span-level scores separately, and the prediction of spans was implemented by voting, as shown in Section~\ref{sec:ident}.

The F1 score for token-level is \textbf{0.985}, while the F1 score for span-level is \textbf{0.988}. The average inference time per page for the classifier is \textbf{0.03} seconds. 

\subsection{PDF-to-Markdown Conversion Speed}
\begin{table}[]
\centering
\caption{\label{tab} The speed of PDF-to-Markdown conversion models and Assisted Generation methods was evaluated across three test datasets, with $t(s)$  representing the average generation time per page in seconds. The extra 0.03 seconds required for CLD is due to the overhead of Copyable Text Identification.}
\begin{adjustbox}{center, width=0.8\linewidth}
\begin{tabular}{lccccccc}
\hline
& & \multicolumn{6}{c}{Test Dataset} \\ \cline{3-8} 
&& \multicolumn{2}{c}{arXiv}& \multicolumn{2}{c}{Economics}& \multicolumn{2}{c}{Quantum Physics}   \\ \hline
\multicolumn{1}{l|}{Backbone Model}& \multicolumn{1}{c|}{Assisted Generation} & $t(s)$& Speed Up& $t(s)$& Speed Up& $t(s)$& Speed Up\\ \hline
\multicolumn{1}{l|}{\multirow{3}{*}{\begin{tabular}[c]{@{}l@{}}Nougat-base\\ (349M)\end{tabular}}}& \multicolumn{1}{c|}{-}& 6.48& -& 5.73& -& 6.80& -\\
\multicolumn{1}{l|}{}& \multicolumn{1}{c|}{mPLD}& 5.92& 1.09$\times$& 5.05& 1.13$\times$& 6.13& 1.11$\times$\\
\multicolumn{1}{l|}{}& \multicolumn{1}{c|}{CLD}& 5.85 (+0.03)  & \textbf{1.10$\times$} & 4.60 (+0.03)  & \textbf{1.24$\times$} & 5.56 (+0.03)  & \textbf{1.22$\times$} \\ \hline
\multicolumn{1}{l|}{\multirow{3}{*}{\begin{tabular}[c]{@{}l@{}}Kosmos-2.5\\ (1.37B)\end{tabular}}}& \multicolumn{1}{c|}{-}& 12.21& -& 11.09& -& 12.77& -\\
\multicolumn{1}{l|}{}& \multicolumn{1}{c|}{mPLD}& 9.42& 1.30$\times$& 8.04& 1.38$\times$& 9.23& 1.38$\times$\\
\multicolumn{1}{l|}{}& \multicolumn{1}{c|}{CLD}& 8.37 (+0.03)  & \textbf{1.45$\times$} & 7.60 (+0.03)  & \textbf{1.45$\times$} & 7.85 (+0.03)  & \textbf{1.62$\times$} \\ \hline
\multicolumn{1}{l|}{\multirow{3}{*}{\begin{tabular}[c]{@{}l@{}}Llama-3.2-Vision\\ (10.6B)\end{tabular}}} & \multicolumn{1}{c|}{-}& 33.83& -& 30.53& -& 35.03& -\\
\multicolumn{1}{l|}{}& \multicolumn{1}{c|}{mPLD}& 29.20& 1.16$\times$& 23.28& 1.31$\times$& 32.08& 1.09$\times$\\
\multicolumn{1}{l|}{}& \multicolumn{1}{c|}{CLD}& 26.02 (+0.03) & \textbf{1.30$\times$} & 21.83 (+0.03) & \textbf{1.40$\times$} & 27.61 (+0.03) & \textbf{1.27$\times$} \\ \hline
\multicolumn{1}{l|}{\multirow{3}{*}{\begin{tabular}[c]{@{}l@{}}Qwen2-VL\\ (12.6B)\end{tabular}}}& \multicolumn{1}{c|}{-}& 138.85& -& 119.42& -& 156.06& -\\
\multicolumn{1}{l|}{}& \multicolumn{1}{c|}{mPLD}& 97.65& 1.32$\times$& 75.21& 1.59$\times$& 100.00& 1.56$\times$\\
\multicolumn{1}{l|}{}& \multicolumn{1}{c|}{CLD}& 91.53 (+0.03) & \textbf{1.41$\times$} & 72.88 (+0.03) & \textbf{1.64$\times$} & 91.88 (+0.03) & \textbf{1.70$\times$} \\ \hline
\end{tabular}
\end{adjustbox}
\end{table}

\begin{figure}
    \centering
    \includegraphics[width=0.8\linewidth]{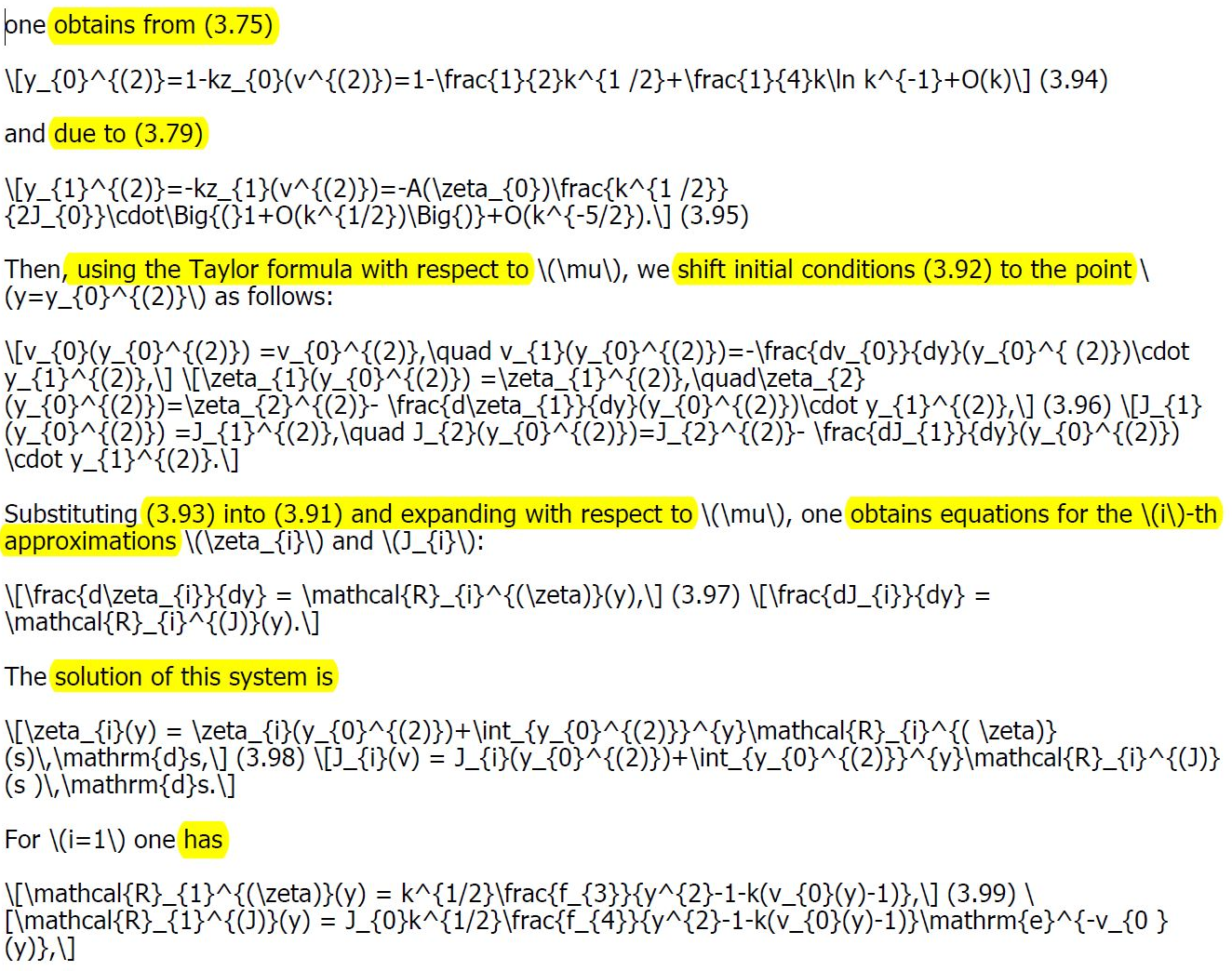}
    \caption{Example of PDF-to-Markdown conversion done by Kosmos-2.5 with the assistance of CLD. In this example, the highlighted parts are the tokens proposed by CLD and accepted by Kosmos-2.5's validation steps.}
    \label{fig:kosmos-example}
\end{figure}

The conversion speed test results are shown in Table~\ref{tab}. In our experiments, both mPLD and CLD speed up PDF-to-Markdown conversion. Larger models benefits more from mPLD and CLD in terms of conversion speed. The inference time of Copyable Text Identification module in CLD is included in the average conversion time per page.
In Figure~\ref{fig:kosmos-example} we show an example of how CLD can accelerate Kosmos-2.5. 

\subsection{Ablation Study}
\begin{table}[]
\centering
\caption{Ablation study results showing the impact of Copyable Text Identification (CTI) and Candidate Generation (CG) on inference time ($t$ in seconds). Both modules contribute to reducing inference time, with their combination achieving the highest speed-up. CLD degrades to the original mPLD method when CTI and CG are removed.}
\label{ablation}
\begin{adjustbox}{center, width=0.8\linewidth}
\begin{tabular}{lcccccccc}
\hline
\multicolumn{1}{c}{} & \multicolumn{8}{c}{Backbone Model} \\ \cline{2-9} 
\multicolumn{1}{c}{} & \multicolumn{2}{c}{Nougat-base} & \multicolumn{2}{c}{Kosmos-2.5} & \multicolumn{2}{c}{Llama-3.2-Vision} & \multicolumn{2}{c}{Qwen2-VL} \\ \hline
\multicolumn{1}{l|}{Modules} & $t(s)$ & Speed Up & $t(s)$ & Speed Up & $t(s)$ & Speed Up & $t(s)$ & Speed Up \\ \hline
\multicolumn{1}{l|}{base (mPLD)} & 6.13 & - & 9.23 & - & 32.08 & - & 100.00 & - \\
\multicolumn{1}{l|}{base + CTI} & 5.80 (+0.03) & 1.05$\times$ & 8.66 (+0.03) & 1.06$\times$ & 27.83 (+0.03) & 1.15$\times$ & 92.83 (+0.03) & 1.07$\times$ \\
\multicolumn{1}{l|}{base + CG} & 5.78 & 1.06$\times$ & 8.68 & 1.06$\times$ & 29.30 & 1.09$\times$ & 96.99 & 1.03$\times$ \\
\multicolumn{1}{l|}{base + CTI + CG} & \textbf{5.60 (+0.03)} & \textbf{1.10$\times$} & \textbf{7.85 (+0.03)} & \textbf{1.17$\times$} & \textbf{27.61 (+0.03)} & \textbf{1.16$\times$} & \textbf{91.88 (+0.03)} & \textbf{1.08$\times$} \\ \hline
\end{tabular}%
\end{adjustbox}
\end{table}

\begin{table}[]
\centering
\caption{Assisted Generation methods losslessly preserve the original output.}
\label{tab:lossless}
\begin{tabular}{lcc}
\hline
\multicolumn{3}{c}{Assisted Generation Quality Loss} \\ \hline
\multicolumn{1}{l|}{Model} & mPLD & CLD \\ \hline
\multicolumn{1}{l|}{Nougat-base} & 0\% & 0\% \\
\multicolumn{1}{l|}{Kosmos-2.5} & 0\% & 0\% \\
\multicolumn{1}{l|}{Llama3.2-Vision} & 0\% & 0\% \\
\multicolumn{1}{l|}{Qwen2-VL} & 0\% & 0\% \\ \hline
\end{tabular}%
\label{quality}
\end{table}

Table~\ref{ablation} demonstrates the impact of different module combinations on inference time and relative speed-up for Nougat-base and Kosmos-2.5 backbone models.
Additionally, ablation studies on Llama and Qwen2-VL show the same trend. 
These findings emphasize the practical relevance of our methods.

\subsection{PDF-to-Markdown Conversion Quality}
\label{sec:quality}
In addition to the conversion speed, we also discuss the conversion quality of the models.
In this work, comparing the accuracy of individual models is highly unfeasible, mainly due to the following reasons:
\begin{itemize}
    \item Assisted Generation is a lossless method and it doesn't change the original output. Therefore, we may quote its accuracy in the model's original papers.
    \item Neither Nougat nor Kosmos-2.5 have publicly released their test datasets, which may be due to copyright reasons described in Section~\ref{sec:dataset}. It is difficult for us to reproduce their results fairly.
    \item The output style of the PDF-to-Markdown conversion task is flexible. Specifically, the reading order of floating elements in a page, including tables and figures, is not fixed; Nougat tends to add floating elements at the end of a page, while Kosmos-2.5 tends to add floating elements at the beginning of a page.
    \item The formatting of tables and formulas is also flexible. Specifically, Nougat uses the \LaTeX{} format for tables, while Kosmos-2.5 uses the Markdown format. Formulas can start and end with either a dollar sign or a backslash with parentheses.
\end{itemize}

For these reasons, we can only provide a very unfair comparison of conversion quality.
Unquestionably, Assisted Generation does not degrade the quality of model generation, as shown in Table~\ref{tab:lossless}.
We adopted the evaluation metric from the Nougat\cite{blecher2023nougat}.
Our evaluation of the four models on our test datasets, as shown in Table~\ref{tab:quality}. 
However, lower scores do not necessarily indicate poorer readability on a manual inspection basis, making this evaluation an imperfect quality assessment. 

\begin{table}[]
\centering
\caption{PDF-to-Markdown Conversion Quality of Different Models. This is not a fair comparison, we are simply presenting the results.}
\label{tab:quality}
\begin{tabular}{l|ccc}
\hline
Models & Edit Dist↓ & BLEU↑ & F1↑ \\ \hline
Nougat-base & 0.38 & 0.54 & 0.73 \\
Kosmos-2.5 & 0.43 & 0.51 & 0.68 \\
Llama-3.2-Vision & 0.53 & 0.42 & 0.61 \\
Qwen2-VL & 0.46 & 0.47 & 0.64 \\ \hline
Marker (Pipeline-based) & 0.45 & 0.44  & 0.62 \\ \hline
\end{tabular}
\end{table}

A case-by-case analysis revealed that Llama and Qwen2-VL struggled to follow instructions precisely. Llama frequently generates incorrect \LaTeX{} code in formulas, often rendering them invalid in Markdown. Additionally, it tends to reconstruct PDF content as \LaTeX{} rather than converting it to Markdown. Both Qwen2-VL and Llama misplace footnotes, placing them at the end of the document instead of their cited positions, increasing their error rates. We plan to refine our evaluation metrics in future work.

\section{Conclusion}
In this paper, we introduced mPLD and CLD, which are lightweight Assisted Generation methods for end-to-end PDF-to-Markdown conversion using decoder transformer models.
mPLD does not add any parameters that require training, and it can be seamlessly added to an existing transformer decoder model.
CLD is based on the off-the-shelf ERNIE Layout model. 
Minimal fine-tuning was performed, and the weights file size is kept at 27 MB.

The ever-increasing number of PDF publications appearing year after year shows no signs of slowing down.
While pipeline-based approaches can help to get the flood of scientific information and new knowledge under control, the development of such technologies is very complex in practice and hinders the creation of infrastructures and systems to track research and assist the scientific community with applications such as dedicated scientific search engines and recommender systems.

Through our efforts, we hope to make end-to-end PDF-to-Markdown conversions with simpler structures that better drive accessibility and structured extraction of scientific information.

\section{Limitations}
In the work, we observed that CLD improves over mPLD mainly by extracting candidates more accurately through a filtering mechanism and a topping mechanism that better recognizes the current PDF-to-Markdown conversion progress. 

However, CLD does not yet explicitly detect the position of the conversion progress of the decoder transformer on the page. The contemporaneous work, LOCR, which predicts the position of the next token on the PDF page using a miniature Prompt Module attached to Nougat, might enhance candidate generation. 
We plan to incorporate LOCR with CLD in the next step.

Kosmos-2.5, Nougat, Llama, and Qwen2-VL discard figures from pages, as they can only generate text.
ERNIE Layout is capable of identifying them. 
Our goal is to explore additional ways to properly integrate figures into Markdown output. For example, Mistral OCR encodes images with base64 strings.

\section*{Acknowledgments}
This work was conducted within the research project InsightsNet (\url{insightsnet.org}), which is funded by the Federal Ministry of Education and Research (BMBF) under grant no. 01UG2130A.
We gratefully acknowledge support from Dr. Wolfgang Stille and the hessian.AI Service Center (funded by the Federal Ministry of Education and Research, BMBF, grant no. 01IS22091) and the hessian.AI Innovation Lab (funded by the Hessian Ministry for Digital Strategy and Innovation, grant no. S-DIW04/0013/003).
We want to thank Dr. Sabine Bartsch for supervising this research.
We also thank the anonymous reviewers for their insightful comments and suggestions, which greatly improved the manuscript.
\bibliographystyle{splncs04}
\bibliography{custom, anthology}
\end{document}